\documentclass[preprint,12pt,authoryear]{elsarticle}

\usepackage{amsmath,amsfonts}
\usepackage{algorithmic}
\usepackage{algorithm}
\usepackage{array}
\usepackage{textcomp}
\usepackage{stfloats}
\usepackage{url}
\usepackage{verbatim}
\usepackage{graphicx}
\usepackage{cite}
\usepackage{hyperref}
\usepackage{multirow}
\usepackage{multicol}
\usepackage{makecell}
\usepackage{diagbox}
\usepackage{balance}
\usepackage{color}
\usepackage{bbding}
\usepackage{ctable} 
\usepackage{tabu}
\usepackage{amssymb}
\usepackage{pifont}
\usepackage{subcaption}
\usepackage{setspace}

\DeclareMathOperator*{\argmax}{arg\,max}

\journal{Computer Speech \& Language}
\begin{document}

\begin{frontmatter}



\title{Elevating Robust Multi-Talker ASR by Decoupling Speaker Separation and Speech Recognition} 


\author{Yufeng Yang$^a$, Hassan Taherian$^a$, Vahid Ahmadi Kalkhorani$^a$, and DeLiang Wang$^{a, b}$} 

\affiliation{organization={Department of Computer Science and Engineering, The Ohio State University},
            addressline={2015 Neil Avenue}, 
            city={Columbus},
            postcode={43210}, 
            state={OH},
            country={United States}}
\affiliation{organization={Center for Cognitive and Brain Sciences, The Ohio State University},
            addressline={1835 Neil Ave}, 
            city={Columbus},
            postcode={43210}, 
            state={OH},
            country={United States}}
            
\begin{abstract}
Despite the tremendous success of automatic speech recognition (ASR) with the introduction of deep learning, its performance is still unsatisfactory in many real-world multi-talker scenarios. Speaker separation excels in separating individual talkers but, as a frontend, it introduces processing artifacts that degrade the ASR backend trained on clean speech. As a result, mainstream robust ASR systems train the backend on noisy speech to avoid processing artifacts. In this work, we propose to decouple the training of the speaker separation frontend and the ASR backend, with the latter trained on clean speech only. Our decoupled system achieves 5.1\% word error rates (WER) on the Libri2Mix dev/test sets, significantly outperforming other multi-talker ASR baselines. Its effectiveness is also demonstrated with the state-of-the-art 7.60\%/5.74\% WERs on 1-ch and 6-ch SMS-WSJ. Furthermore, on recorded LibriCSS, we achieve the speaker-attributed WER of 2.92\%. These state-of-the-art results suggest that decoupling speaker separation and recognition is an effective approach to elevate robust multi-talker ASR.
\end{abstract}




\begin{keyword}
LibriCSS \sep Libri2Mix \sep robust multi-talker ASR \sep SMS-WSJ \sep speaker separation \sep speech distortion 



\end{keyword}

\end{frontmatter}



\section{Introduction}\label{sec:intro}
In the era of deep learning, automatic speech recognition (ASR) has made tremendous strides, progressing from conventional Gaussian mixture model plus hidden Markov model (GMM-HMM) approaches \citep{rabiner1989tutorial}, to hybrid systems of deep neural network and HMM (DNN-HMM) \citep{hinton2012deep} to end-to-end (E2E) systems \citep{prabhavalkar2023end}. ASR has been seamlessly integrated into our daily lives as an indispensable component in personal assistants and home devices, significantly boosting human-computer interaction. Despite its success, ASR is prone to acoustic interference, such as reverberation, background noise, and overlapping speakers. The mismatch between speech signals in such environments and the transcribed training data in clean conditions pose a persistent obstacle, demanding the development of robust ASR systems capable of overcoming this mismatch \citep{vincent2013second, vincent2017analysis}.

Meanwhile, the introduction of deep learning has also dramatically improved speech separation performance \citep{wang2018supervised}, including the intelligibility and quality of degraded speech signals. To address the robust ASR problem, a straightforward approach is to employ a speech separation model as the frontend for an ASR backend model. Speech separation models operate in the time \citep{luo2019conv, pandey2022self} or time-frequency (T-F) domains \citep{tfgridnet_2023_wang, quan2023spatialnet, vahid2024crossnet} for speech enhancement (speech-noise separation) or multi-talker speaker separation. Despite these effective frontends, the processing artifacts introduced in separation can be detrimental to the ASR backend trained on clean speech \citep{wang_bridging_2019}.

To address this challenge, prevailing approaches train an ASR model directly on noisy speech \citep{vincent2013second, vincent2017analysis, yang2022conformer} or enhanced speech \citep{wang_bridging_2019}, or train a joint system of speech separation frontend and ASR backend \citep{chang2022end}. In multi-talker or conversational cases, training ASR on overlapped speech directly is problematic due to the large variety of mixed talkers. Although there are multi-talker ASR methods, including serialized output training~\citep{kanda2020serialized} and other techniques~\citep{guo2021multi, zhang2023weakly, zhang2023exploring, shi2024serialized, shi2024advancing, shi2024keyword}, their performance degrades on single-talker ASR. Also, the amount of annotated data for conversational ASR is much smaller than that for single-talker ASR. Hence, speaker separation becomes attractive to address overlapped speech for robust ASR \citep{qian2018single}, and permutation invariant training (PIT) is employed in many related studies \citep{kolbaek2017multitalker, chang2020mtasr, lu2021streaming}. The training target of a speaker separation frontend is usually clean speech. Thus, the choice of training data for the ASR backend becomes critical to achieving high performance. The mainstream strategy trains ASR on single-talker speech with various noise augmentations as done, for example, in the baseline system of the SMS-WSJ corpus \citep{drude2019sms}. However, when tested on clean speech, a performance gap arises between the ASR model trained on noisy speech and that trained on clean speech. By clean speech, we include single-talker utterances recorded in a variety of quiet places in the real environment, as done in the collection of the LibriSpeech corpus~\citep{panayotov2015librispeech}. As speech separation performance continues to improve, is training the ASR backend on noisy speech still the most effective approach, given the mismatch between the separated speech that is aimed to be clean and the noisy training speech?

In this work, we aim to address such mismatch and elevate robust ASR performance. We investigate robust ASR in multi-talker scenarios and propose to decouple the two stages of speaker separation and speech recognition. Our approach separates the training of a speaker separation frontend and an ASR backend trained on clean speech only, i.e., each stage is trained separately without considering the other. In this way, the mismatch between the frontend output and the backend training data is expected to be alleviated. As shown in \citet{yang2022time, yang2024towards}, with monaural speech enhancement, ASR trained on clean speech outperforms that trained on noisy speech. We do not investigate a jointly trained model as it has been shown that, after joint fine-tuning, the performance of each part degrades on its original task for a robust ASR system designed with a speaker separation frontend and ASR backend~\citep{masuyama2023exploring}.

The proposed decoupled approach is evaluated on the Libri2Mix \citep{cosentino2020librimix}, SMS-WSJ \citep{drude2019sms}, an extension of SMS-WSJ with a larger size called SMS-WSJ-Large, and LibriCSS \citep{chen2020libricss} corpora. On Libri2Mix, we achieve the state-of-the-art word error rate (WER) of 5.1\% on the dev/test sets, significantly outperforming other multi-talker ASR baselines. By training the backend on clean speech only, we achieve 7.60\% and 5.74\% WER on 1-ch and 6-ch SMS-WSJ test sets, outperforming the previous best that trains the backend on reverberant-noisy speech with three times our training data \citep{tfgridnet_2023_wang, quan2023spatialnet}. Moreover, on recorded LibriCSS, we obtain the speaker-attributed WER of $2.92\%$, outperforming the previous best \citep{taherian2023ssnd} by $9.3\%$ with an ASR backend trained on clean speech. We also evaluate the generalization ability of the decoupled system with different training data on SMS-WSJ-Large in 1-ch and 6-ch conditions. These results with different frontend and backend combinations show that the proposed decoupled approach can substantially elevate the performance of robust ASR, and the ASR backend does not have to be trained on noisy speech as done in the mainstream approach.

The main contributions are summarized as follows. First, we develop a decoupled robust ASR system where the speaker separation frontend and ASR backend are independently trained, with the backend trained on clean speech only. Second, this work demonstrates that, with a powerful speaker separation frontend, training ASR on clean speech can elevate recognition performance, as shown in our advancement of state-of-the-art results on the Libri2Mix, SMS-WSJ, and LibriCSS datasets. Hence, we believe that the decoupled approach offers a strong alternative to the mainstream approach for robust multi-talker ASR.

This paper expands a preliminary study~\citep{yang2025elevating_conf} in several major ways. First, we perform evaluations on more corpora, including Libri2Mix, SMS-WSJ, and SMS-WSJ-Large. Second, in addition to the analysis in multi-channel conditions, we evaluate on single-channel Libri2Mix, SMS-WSJ, and SMS-WSJ-Large. Third, we add another ASR backend on LibriCSS that is trained with self-supervised learning (SSL) features~\citep{chen2022wavlm}. Fourth, we introduce a new dataset SMS-WSJ-Large, which is easy to generate and reproduce. The evaluation on this dataset indicates that the effective decoupling is not limited to multi-channel conditions for robust multi-talker ASR, and provides insight into the conditions where the proposed approach outperforms the mainstream approach.

The remainder of the paper is organized as follows. Section~\ref{sec:system} presents the system model and describes the DNN architectures of the frontend and backend. Section~\ref{sec:exp} describes the experimental setup and implementation details. Section~\ref{sec:result} presents the evaluation results and comparisons. Conclusions and discussions are made in Section~\ref{sec:conclusion}.

\section{Methods}\label{sec:system}
\subsection{Problem Formulation}
\subsubsection{Speaker Separation}
The physical model of a $P$-microphone mixture can be formulated in the short-time Fourier transform (STFT) domain as

\begin{equation}
\begin{aligned}
      \mathbf{Y}(t, f) & = \sum_{c=1}^{C} \mathbf{X}(c, t, f) + \mathbf{N}(t, f)\\
    & =  \sum_{c=1}^{C} (\mathbf{S}(c, t, f) + \mathbf{H}(c, t, f)) + \mathbf{N}(t, f),
\end{aligned}
\end{equation}
where $C$ is the number of speakers, which is set to $2$ in this study, assuming at most 2 speakers talk simultaneously. $\mathbf{Y}(t, f)$, $\mathbf{N}(t, f)$, $\mathbf{X}(c, t, f)$, $\mathbf{S}(c, t, f)$, and $\mathbf{H}(c, t, f)$ $\in \mathbb{C}^{P}$ respectively denote the STFT of the received mixture, reverberant noise, reverberant speech, direct-path signal and reflections of speaker $c$ at time $t$ and frequency $f$ . We drop the index of $t$ and $f$ in later notations. Speaker separation aims to estimate $S_q(c)$ for each source at a reference microphone $q$ given input $\mathbf{Y}$. 

\subsubsection{Automatic Speech Recognition}
An ASR system estimates a word sequence $\mathbf{W^{*}}$ given a sequence of acoustic features $\mathbf{A}$ of speech signal $\mathbf{a}$, which can be formulated as 

\begin{equation}\label{eq:map}
    \mathbf{W^{*}} = \argmax_{\mathbf{W}} P_{\mathcal{AM}, \mathcal{LM}}(\mathbf{W} | \mathbf{A}),
\end{equation}
where $\mathcal{AM}$ and $\mathcal{LM}$ denote an acoustic model (AM) and language model (LM), respectively. Using Bayes' theorem, Equation~\ref{eq:map} can be written as

\begin{equation}
    \mathbf{W^{*}} = \argmax_{\mathbf{W}} p_{\mathcal{AM}}(\mathbf{A} | \mathbf{W})P_{\mathcal{LM}}(\mathbf{W}),
\end{equation}
where $p_{\mathcal{AM}}$ and $P_{\mathcal{LM}}$ are AM likelihood and LM prior probability, respectively. The AM predicts the likelihood of acoustic features of a phoneme or another linguistic unit, and the LM provides a probability distribution over words or sequences of words in a speech corpus. In an E2E ASR system, the word sequence is predicted directly given $\mathbf{A}$.

\subsection{Speaker Separation Frontend}
\subsubsection{SpatialNet}\label{sec:frontend_spatialnet}
SpatialNet~\citep{quan2023spatialnet} is employed as a T-F domain multi-channel speaker separation frontend. It has interleaved narrow-band and cross-band blocks to exploit narrow-band and across-frequency spatial information, respectively. The narrow-band blocks process frequencies independently, and use a self-attention mechanism and temporal convolutional layers to perform spatial-feature-based speaker clustering and temporal smoothing and filtering, respectively. The cross-band blocks process frames independently, and use full-band linear layer and frequency convolutional layers to learn the correlation among all frequencies and adjacent frequencies, respectively.

SpatialNet performs multi-channel complex spectral mapping~\citep{wang2020multi} by predicting the real and imaginary (RI) parts of the STFT of each talker from the stacked RI parts of the STFT of overlapped speech~\citep{williamson2015complex}. The separated waveforms are generated by performing an inverse STFT on the estimated RI parts.

\subsubsection{TF-CrossNet}
We use TF-CrossNet as another T-F domain multi-channel speaker separation frontend \citep{vahid2024crossnet}. Different from SpatialNet, TF-CrossNet introduces positional encoding and a cross-temporal module after the cross-band module. This modification enhances temporal processing. TF-CrossNet also performs complex spectral mapping for speaker separation.

\subsubsection{Speaker Separation via Neural Diarization}
We employ speaker separation via neural diarization (SSND) \citep{taherian2023ssnd} for multi-talker speech recognition \citep{chen2020libricss}. SSND performs speaker separation by integrating speaker diarization. The diarization is performed with multi-channel end-to-end neural diarization through an encoder-decoder-based attractor module (MC-EEND), which is trained with location-based training (LBT) \citep{taherian2022multi} to resolve the permutation ambiguity issue in talker-independent speaker separation. After diarization, a sequence of speaker embeddings computed from non-overlapped speech frames is used to facilitate the assignment of speakers to the output streams of the speaker separation model. In this way, speaker assignment is accomplished during the diarization process, instead of the speaker separation process.

\subsection{Speech Recognition Backend}
\subsubsection{Factorized Time-delayed Neural Network}
We utilize a factorized time-delayed neural network (TDNN-F) based on the WSJ Kaldi recipe~\citep{povey2011kaldi} as one ASR backend. The AM consists of 8 TDNN-F layers, and the final
word sequence is obtained by decoding the state posteriors with a default WSJ tri-gram Kaldi language model without additional
N-best restoring. More details can be found in~\citet{drude2019sms}.

\subsubsection{Wide-residual Conformer}
We utilize an E2E ASR model in \citet{yang2024towards}, which is a connectionist temporal classification (CTC) and attention Conformer-encoder Transformer-decoder E2E ASR model, denoted as wide-residual Conformer (WRConformer)~\footnote{\url{https://github.com/yfyangseu/espnet}}. It leverages the ASR recipe in ESPnet \citep{watanabe2018espnet}, and adapts the standard CTC and attention Conformer encoder Transformer decoder ASR recipe to WRConformer. In this adaptation, the 2-D convolution in the subsampling module is replaced by a modified wide-residual convolutional neural network (WRCNN), which comprises two ResBlocks (see \citet{jahn2016wide}). The first ResBlock projects an input log-Mel feature to 512 dimensions, while the second Resblock maintains the same input and output dimensions. Each ResBlock subsamples time frames by a factor of 2, resulting in the total number of frames reduced by a factor of 4 after the subsampling module, matching the default ESPnet subsampling module. In WRConformer, the number of Conformer encoders is set to 10 and the other configurations are the same as the default ESPnet setting.

\subsection{Proposed Decoupled Approach}\label{sec:decoupled_approach}
\begin{figure}[!htbp]
    \centering
    \includegraphics[width=0.8\linewidth]{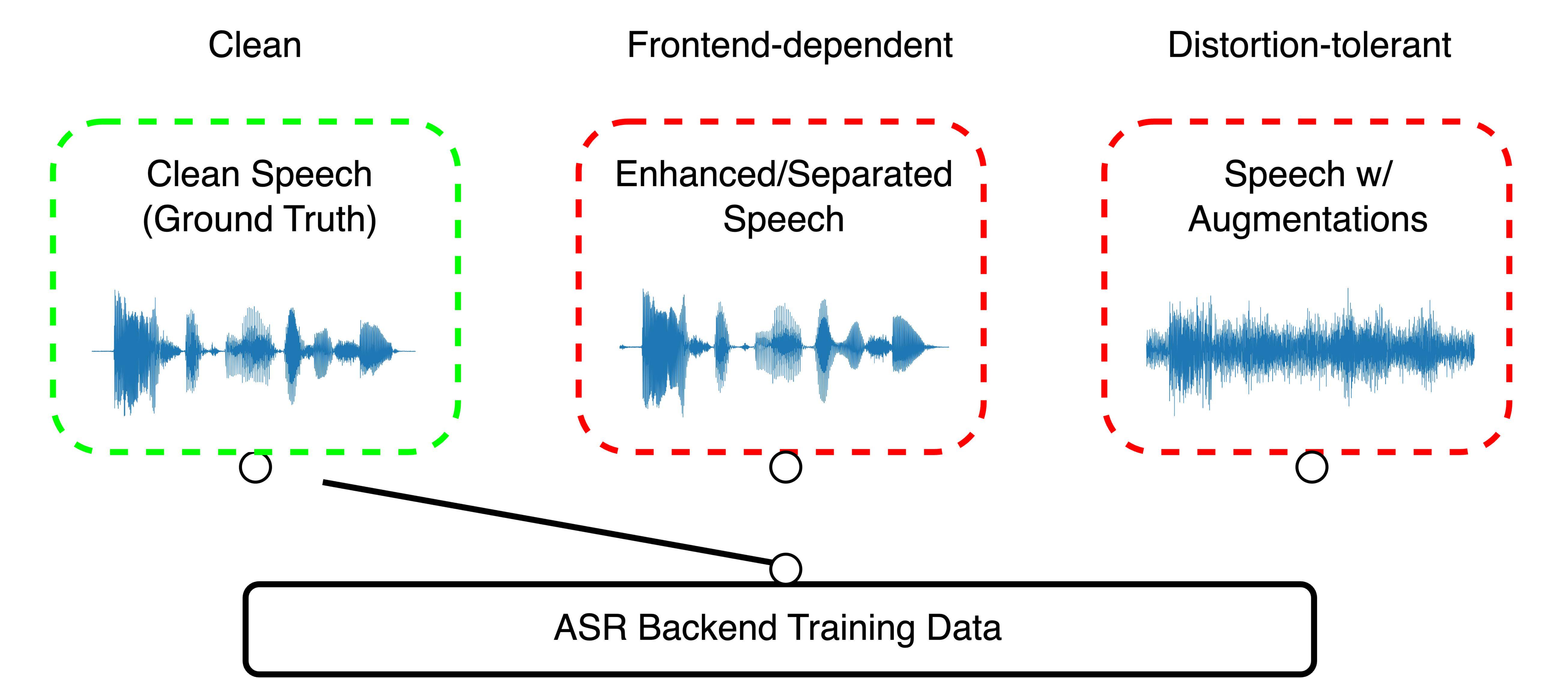}
    \caption{Comparison of different ASR backend training approaches.}
    \label{fig:decoupled}
\end{figure}

We introduce the decoupled approach by presenting a comparison of different ASR backend training data in Figure~\ref{fig:decoupled}. Training the ASR backend on single-talker speech with reverberation and noise augmentations aims to mitigate acoustic distortions and processing artifacts, which is the mainstream approach and is denoted as distortion-tolerant. However, it yields suboptimal results when tested on clean speech, as discussed earlier. On the other hand, the ASR backend trained on enhanced or separated speech is frontend-dependent, making the system inflexible. Retraining is needed when a better frontend or backend needs to be incorporated to achieve improved performance. To address these limitations and leverage the capabilities of powerful separation frontends, we propose to train the ASR backend on clean speech only. The proposed decoupled system significantly improves the flexibility compared to the frontend-dependent approach, and elevates the performance on clean speech relative to the distortion-tolerant approach.

\section{Experimental Setup}\label{sec:exp}
\subsection{Datasets}
\subsubsection{Libri2Mix}
We use the ESPnet version\footnote{\url{https://github.com/espnet/espnet/tree/master/egs2/librimix/sot_asr1}} of Libri2Mix~\citep{cosentino2020librimix}, which is designed specifically for multi-talker ASR. The data generation process is based on the official implementation\footnote{\url{https://github.com/JorisCos/LibriMix}} for speaker separation and an offset is added to one speaker in each speech mixture. The dataset utilizes clean speech sourced from LibriSpeech~\citep{panayotov2015librispeech} and noise from WHAM!~\citep{wichern2019wham} datasets. The training set has 50800 and 13900 utterances in train-360 and train-100 sets, respectively. The development set and test set have 3000 utterances each. The sampling rate is 16 kHz. In this paper, we focus on the challenging robust multi-talker ASR in noisy conditions for Libri2Mix.

\subsubsection{SMS-WSJ and SMS-WSJ-Large}
We employ SMS-WSJ \citep{drude2019sms} to evaluate ASR with multi-channel speaker separation in reverberant conditions. The dataset consists of 33561, 982, and 1332 train, validation, and test mixtures, respectively. All utterances are drawn from the WSJ0 and WSJ1 datasets \citep{paul1992wsj}. The sampling rate is 8 kHz and the longer utterance determines the length of the mixture. The sensor array is a circle with a radius of 10 cm. T60s are sampled from [0.2, 0.5] s, and the distance between the array and the speaker ranges in [1, 2] m. Additive white sensor noise is added with signal-to-noise ratio (SNR) ranges in [20, 30] dB. The first microphone is selected as the reference microphone.

To evaluate the generalization ability and the effect of training data size on a robust multi-talker ASR system, we create a new dataset called SMS-WSJ-Large, by modifying the data generation process of SMS-WSJ\footnote{\url{https://github.com/fgnt/sms_wsj}}. We multiply the original training data size by a factor of 4, resulting in 134244 training utterances. The numbers of utterances in the development and test sets are kept the same as SMS-WSJ. The training and development data T60 and SNR ranges are modified to [0.2, 1.1] s and [0, 40] dB, respectively. For the test set, we consider three T60 ranges: [0.2, 0.5], [0.5, 0.8], and [0.8, 1.1] s, and four SNR ranges: [0, 10], [10, 20], [20, 30], and [30, 40] dB. By combining each T60 and SNR range, 12 test sets are generated for evaluation in SMS-WSJ-Large. The random seed is the same as in \citet{drude2019sms} for easy replication.

\subsubsection{LibriCSS}
LibriCSS is a dataset designed for evaluating multi-talker speech recognition \citep{chen2020libricss}. The dataset has 10 one-hour sessions, each of which has 6 ten-minute mini-sessions with different overlap levels. These levels are 0S (no speaker overlap and short inter-utterance silence ranges in [0.1, 0.5] s), 0L (no speaker overlap and long inter-utterance silence ranges in [2.9, 3.0] s), and speaker overlap ratio at $10\%$, $20\%$, $30\%$, and $40\%$. The utterances are drawn from the LibriSpeech test-clean set \citep{panayotov2015librispeech} and the sampling rate is 16 kHz. The recordings are made of loudspeakers in a real meeting room with a seven-channel circular microphone array, which has six microphones evenly placed on a circle with a radius of 4.25 cm and an additional center microphone.

The meeting-style training data for MC-EEND and SSND is generated following a LibriCSS recipe\footnote{\url{https://github.com/jsalt2020-asrdiar/jsalt2020_simulate}}, where the LibriSpeech training set is utilized. We follow the data generation setup of \citet{taherian2023ssnd} for speaker diarization and speaker separation training. The center microphone is treated as the reference microphone.

\subsection{Frontend Configurations}
\subsubsection{Libri2Mix}\label{sec:exp_libri2mix}
We employ TF-CrossNet \citep{vahid2024crossnet} for 1-ch speaker separation on Libri2Mix. The network is configured with 12 TF-CrossNet blocks where the number of hidden channels is set to 192 and the cross-band hidden dimension is 16. In addition, the narrow-band hidden dimension is set to 384, and the number of attention heads is 4. The STFT frame size and shift are 16 and 8 ms, respectively. Additionally, random chunk positional encoding is employed for training. A combination of RI-Mag loss \citep{wang2020complex} and the scale-invariant signal-to-distortion ratio (SI-SDR) loss \citep{leroux2019sdr} is utilized for model training. These loss functions are defined as

\begin{subequations}
    \begin{align}
        \mathcal{L} &= \mathcal{L}_{\text{RI-Mag}}+\mathcal{L}_{\text{SI-SDR}},\\
        \mathcal{L}_\text{RI-Mag} & = \Big\Vert \mathbf{S}_{r} - \mathbf{\hat{S}}_r  \Big\Vert_{1} + \Big\Vert \mathbf{S}_{i} - \mathbf{\hat{S}}_i  \Big\Vert_{1} + \Big\Vert |\mathbf{S} - \mathbf{\hat{S}} |  \Big\Vert_{1},\\
    \mathcal{L}_{\text{SI-SDR}} &= -\sum_{c=1}^{C}10\log_{10}\frac{\Vert \mathbf{s}_{c}\Vert_{2}^{2}}{\Vert \mathbf{\hat{s}}_{c} - \alpha_{c}\mathbf{s}_{c}\Vert_{2}^{2}},\\
    \alpha_c &= \frac{\mathbf{s}_{c}^{T}\mathbf{\hat{s}}_{c}}{\mathbf{s}_{c}^{T}\mathbf{s}_{c}},
    \end{align}
\end{subequations}
where $\mathbf{s}$ and $\mathbf{\hat{s}}$ denote the ground-truth and estimated speech, and $\mathbf{S}$ and $\mathbf{\hat{S}}$ denote their STFTs, respectively. Subscript $r$ and $i$ denote the real and imaginary parts of STFT, respectively. $|\cdot|$ denotes magnitude, $||\cdot||_1$ denotes the $L_1$ norm, and $||\cdot||_2$ denotes the $L_2$ norm. Term $\alpha_c$ is the scaling factor. The loss function is computed via utterance-level PIT \citep{kolbaek2017multitalker}. The network training is performed on 4 NVIDIA H100 GPUs with 94 GB memory for 200 epochs, employing automatic mixed precision for accelerated training for the first 150 epochs and full precision training for the last 50 epochs. We utilize the Adam optimizer \citep{kingma2015adam} with a maximum learning rate of 0.001. We use a cosine warm-up scheduler that increases the learning rate from $1e^{-4}$ to $1e^{-3}$ in the first 10 epochs. Afterwards, the ReduceLROnPlateau scheduler in PyTorch is utilized, with patience of 3 epochs and a factor of 0.8. The batch size is set to the maximum number that can fit into the GPU memory. The final model with the best validation SI-SDR value is selected for evaluation.

\subsubsection{SMS-WSJ and SMS-WSJ-Large}\label{sec:exp_smswsj}
For SMS-WSJ, we also employ TF-CrossNet and keep the same training configuration as in \citet{vahid2024crossnet} and Section~\ref{sec:exp_libri2mix}. The differences from Section~\ref{sec:exp_libri2mix} are that the STFT frame size and shift are 32 and 16 ms, respectively. Additionally, long short-term memory (LSTM) and random chunk positional encoding are employed for 1-ch and 6-ch training, respectively. The training is performed on 4 NVIDIA A100 GPUs with 80 GB memory.

For training on 6-ch SMS-WSJ-Large, we follow the same procedure as that for 6-ch SMS-WSJ and train a TF-CrossNet model from scratch. For 1-ch SMS-WSJ-Large, we take model weights from the pre-trained model on SMS-WSJ 1-ch, and then fine-tune it with SI-SDR loss $\mathcal{L}_{\text{SI-SDR}}$ on the 1-ch SMS-WSJ-Large data. In this fine-tuning, the learning rate of the Adam optimizer is lowered to $1e^{-5}$, and in the learning rate schedule, the initial and peak learning rate is lowered to $1e^{-5}$ and $1e^{-4}$, respectively. To make fair comparisons, we also fine-tune the same pre-trained model on 1-ch SMS-WSJ with the same loss function and learning rate schedule, and take this fine-tuned model on the same dataset as a pre-trained model in Section~\ref{sec:eval_smswsj_large}.

\subsubsection{LibriCSS}
We employ the same setup as \citet{taherian2023ssnd} for MC-EEND and SSND training. The MC-EEND encoder for diarization utilizes eight Transformer blocks, each with 16 attention heads and a hidden dimension of 256. For speaker separation, for fair comparisons, we use SpatialNet-large \citep{quan2023spatialnet}, consisting of 12 blocks, $D$ = 192 channels, narrowband hidden dimensionality of 384, and cross-band hidden dimensionality of 16. STFT window size and shift are 32 and 16 ms, respectively. SpatialNet incorporates speaker embedding sequences with multi-channel speech mixtures, processed through separate encoders and stacked for subsequent SpatialNet blocks. RI-Mag loss $\mathcal{L}_{\text{RI-Mag}}$ is utilized to train SpatialNet.

\subsection{Backend Configurations}
\subsubsection{Libri2Mix}\label{sec:exp_asr_libri2mix}
We utilize WRConformer as the ASR backend for Libri2Mix. It has 10 Conformer encoders, 6 Transformer decoders, and an attention dimension of 512 with 8 attention heads. The feedforward layer operates with a dimension of 2048. A dropout rate of 0.1 is applied. The CTC weight is set to 0.3, and the label smoothing weight is 0.1. The STFT frame size and shift are 512 and 160, respectively. All WRConformers are trained on WavLM extracted features~\citep{chen2022wavlm} for Libri2Mix.

We train several WRConformers for comparison. For models trained on clean speech, one is trained on the train-clean-100 and train-clean-360 sets of LibriSpeech, and validated on the LibriSpeech dev set, denoted as ``LibriSpeech Clean". Another is trained on single-speaker clean speech of Libri2Mix (s1 and s2 in train-100 and train-360 sets), and validated on the Libri2Mix clean dev set, denoted as ``Libri2Mix Clean".

For distortion-tolerant models trained on noisy speech, one is trained on Libri2Mix single-speaker noisy speech (mix\_single in train-100 and train-360) denoted as ``Libri2Mix Noise-dependent". For another model, we dynamically and randomly mix the noise (noise in train-100 and train-360) and single-speaker clean speech (s1 and s2 in train-100 and train-360) as training data, and the trained model is denoted as ``Libri2Mix Noise-independent". The training SNR is sampled from a normal distribution of $\mathcal{N}(-2, 3.6)$ dB according to \citet{cosentino2020librimix}. We also mix clean speech from LibriSpeech train-100 and train-360 with noise in Libri2Mix train-100 and train-360 dynamically and randomly with the same SNR as in Libri2Mix Noise-independent, denoted as ``LibriSpeech Noise-independent". All distortion-tolerant WRConformers are validated on the Libri2Mix noisy single-speaker dev set to ensure their noise robustness. We also employ a generic robust ASR model of Whisper Large-v3 \citep{whisper} for comparison. During the ASR decoding stage, no LM is used.

\subsubsection{SMS-WSJ and SMS-WSJ-Large}\label{sec:exp_backend_smswsj}
The ASR backend is based on the TDNN-F AM~\citep{drude2019sms}. We train five AMs on different types of data. The task-standard AM is trained on reverberant-noisy speech from the first, third, and fifth microphones. We train four additional AMs on WSJ 8 kHz (denoted as WSJ) clean speech, direct-path speech, 1-ch TF-CrossNet separated training set, and 6-ch TF-CrossNet separated training set, all from the first microphone only. The alignment is based on the WSJ clean speech for the AM trained on WSJ, and on the first channel of direct-path speech for all other AMs. The training and decoding setup for all AMs follow the task-standard settings. The TF-CrossNet-separated training set is used only as a reference of performance upper bound because this will create a dependency between the frontend and backend. For evaluation on SMS-WSJ-Large, we employ the task-standard AM and the AM trained on direct-path speech, representing the mainstream approach and the proposed decoupled approach, respectively.

\subsubsection{LibriCSS}
We utilize WRConformer as the backend for LibriCSS. We train two WRConformers, one is based on log-Mel features (denoted as Our E2E) and the other is based on WavLM extracted features (denoted as Our E2E-SSL). WRConformers are trained on LibriSpeech for 50 epochs on 4 A100 GPUs with the same configuration as the ESPnet LibriSpeech recipe in \citet{watanabe2018espnet}. We compute the concatenated minimum-permutation WER (cpWER) \citep{watanabe2020chime}, which is calculated by concatenating all utterances of each speaker for both reference and hypothesis, then computing the WER between the reference and all possible speaker permutations of the hypothesis, and finally picking the lowest WER value. Tested on the LibriSpeech test-clean/test-other sets, our E2E and our E2E-SSL respectively achieve 1.9\%/4.1\% and 1.8\%/3.5\% WERs.

\section{Results and Discussions}\label{sec:result}
\subsection{Results on Libri2Mix}
The WER results of the proposed approach and the comparison with other systems are presented in Table~\ref{tab:libri2mix_comparison}. We compare the WERs on the development and test set of Libri2Mix with other baselines. For the proposed approach, we also evaluate and compare the WERs of different ASR backends on the Libri2Mix single-speaker test clean speech, denoted as ``Ground Truth" in Table~\ref{tab:libri2mix_comparison}.

For comparison systems, the official ESPnet baseline using serialized output training (SOT) \citep{kanda2020serialized} and Conformer \citep{conformer}, and another baseline utilizing WavLM to extract features, achieve 23.3\% and 17.1\% WER on the test set, respectively. In \citet{guo2021multi}, a non-autoregressive model based on a conditional chain model with Conformer and CTC loss achieves 24.9\% WER. Target-speaker HuBERT (TS-HuBERT) based method \citep{zhang2023weakly} achieves 24.8\% WER on the test set, and target speaker extraction (TSE) based method \citep{zhang2023exploring} achieves 12.0\% WER on the test set. In \citet{shi2024serialized}, overlapped encoding separation is utilized, and the serialized speech information guidance SOT (GEncSep) based method achieves 15.0\% WER on the test set. Large language models (LLMs) are employed in \citet{shi2024advancing} to help multi-talker ASR, and with fine-tuning on Libri2Mix, the best system achieves 9.0\% WER on the test set. To the best of our knowledge, the 7.0\% WER in \citet{shi2024keyword} represents the current best result on the Libri2Mix dataset, and this system utilizes a multi-talker ASR model trained on the speech material of the full LibriSpeech dataset.

\begin{table}[!t]
    \centering
    \caption{ASR ($\%$WER) results of the proposed and comparison systems on Libri2Mix. }
    \label{tab:libri2mix_comparison}
    \centering
    \resizebox{\linewidth}{!}{
    \begin{tabular}[width=\linewidth]{lccc c c}
        \specialrule{1pt}{0pt}{4pt}
         \multirow{2}{*}{Systems} & \multirow{2}{*}{w/ Frontend} & \multirow{2}{*}{w/ SSL} & \multicolumn{3}{c}{WER} \\
         & & & Dev & Test & Ground Truth \\
         \midrule
         SOT-Conformer~\citep{watanabe2018espnet} & \ding{55} & \ding{55} & 24.7 & 23.3 & - \\
         SOT-WavLM-Conformer~\citep{watanabe2018espnet} & \ding{55} & \ding{51} & 19.4 & 17.1 & -\\
         Conditional-Conformer~\citep{guo2021multi} & \ding{55} & \ding{55} & 24.5 & 24.9 & -\\
         TS-HuBERT~\citep{zhang2023weakly} & \ding{55} & \ding{51} & - & 24.8 & -\\
         TSE-Whisper~\citep{zhang2023exploring} & \ding{51} & \ding{55} & - & 12.0 & -\\
         GEncSep~\citep{shi2024serialized} & \ding{55} & \ding{51} & 17.2 & 15.0 & -\\
         WavLM LLM~\citep{shi2024advancing} & \ding{55} & \ding{51} & 11.4 & 10.2 & -\\
         WavLM LLM w/ Fine-tune~\citep{shi2024advancing} & \ding{55} & \ding{51} & 10.3 & 9.0 & -\\
         PIT-SOT~\citep{shi2024keyword} & \ding{55} & \ding{55} & - & 7.0 & -\\
        \midrule
         Libri2Mix Clean & \ding{51} & \ding{51} & 5.1 & 5.1 & 2.3\\
         Libri2Mix Noise-dependent & \ding{51} & \ding{51} & 5.0 & 4.9 & 2.9\\
         Libri2Mix Noise-independent & \ding{51} & \ding{51} & 5.2 & 5.0 & 2.8\\
         LibriSpeech Clean & \ding{51} & \ding{51} & 5.2 & 5.2 & 4.1\\
         LibriSpeech Noise-independent & \ding{51} & \ding{51} & 5.2 & 5.2 & 3.9\\
         Whisper Large-v3~\citep{whisper} & \ding{51} & \ding{55} & 5.6 & 5.0 & 2.5\\
         \specialrule{1pt}{0pt}{0pt}
    \end{tabular}
    }
\end{table}

Now, we analyze and compare the results of systems that follow the proposed decoupled approach with different ASR backends. We first use TF-CrossNet to separate the noisy Libri2Mix two-talker mixtures into two streams of separated speech. Then, with the resulting permutation, WERs are computed given the transcripts of Ground Truth. We report the WERs on the development and test sets with several ASR backends described in Section~\ref{sec:exp_asr_libri2mix}. All ASR backends have similar performances, with the best of 4.9\% from the Libri2Mix Noise-dependent model. When the ASR backend is trained on clean speech only, Libri2Mix Clean and LibriSpeech Clean yield very close 5.1\% and 5.2\% WERs, respectively. Even for the Whisper Large-v3, the 5.0\% WER is the same as the Libri2Mix Noise-independent model. Our systems significantly outperform the current state-of-the-art result of 7.0\% WER by over 25\% relatively, demonstrating the effectiveness of the decoupled approach. 

Since several systems perform similarly on Libri2Mix, which ASR backend should we select? To answer this question, we evaluate each ASR backend on the Ground Truth, and find that Libri2Mix Clean yields the best WER. It outperforms the Libri2Mix Noise-dependent and Libri2Mix Noise-independent models, indicating that training the backend on single-speaker noisy speech degrades the performance on clean speech, as mentioned in Section~\ref{sec:intro} and Section~\ref{sec:decoupled_approach}. Comparing with the ASR backend that yields the best performance on the Libri2Mix test set, i.e., Libri2Mix Noise-dependent, Libri2Mix Clean underperforms by 3.9\% relatively on the Libri2Mix test set but outperforms on Ground Truth by 20.7\%. The second best system on Ground Truth comes from Whisper Large-v3, which underperforms by 2.0\% on the Libri2Mix test set compared to Libri2Mix Noise-dependent, and outperforms on Ground Truth by 13.8\%. Taking the difference of the training data sizes of Libri2Mix Clean and Whisper into consideration, we conclude that the answer to the earlier question is Libri2Mix Clean. These results suggest that, when a powerful speaker separation frontend is available, training the ASR backend on clean speech is an effective option for robust multi-talker ASR, which stands in sharp contrast to the mainstream approach that trains ASR backend on noisy speech.

\subsection{Results on SMS-WSJ}

\begin{table}[!b]
    \caption{Speaker separation results on SMS-WSJ (1-channel).}
    \label{tab:smswsj_comparison_1ch}
    \centering
    \resizebox{0.9\linewidth}{!}{
        \begin{tabular}{@{}lccccc@{}}
            \specialrule{1pt}{0pt}{0pt}
            Model    & SI-SDR   & SDR   & PESQ & eSTOI & WER  \\
            \hline
            Unprocessed    & -5.5   & -0.4   & 1.50 & 0.441 & 79.11  \\
            Oracle direct-path   & $\infty$ & $\infty$ & 4.50 & 1.000 & 6.16 \\
            \hline
            DPRNN-TasNet \citep{luo2020dprnn}  & 6.5  & - & 2.28 & 0.734 & 38.10 \\
            SISO$_{1}$ \citep{wang2021multi}   &   5.7   &  -  & 2.4 & 0.748 &  28.70                \\
            DNN$_{1}$+(FCP+DNN$_{2}\text{)}\times$2 \citep{wang2021multi}  &  12.7  & 14.1        & 3.25 & 0.899 & 12.80   \\
            DNN$_{1}$+(msFCP+DNN$_{2}\text{)}\times$2   \citep{wang2021convolutive}          &   13.4   &  - & 3.41 & - &  10.90      \\
            
            TF-GridNet (1-stage)\citep{tfgridnet_2023_wang} & 16.2     & 17.2     & 3.45 & 0.924 & 9.49                  \\
             TF-GridNet (2-stage)\citep{tfgridnet_2023_wang} & 18.4     & 19.6     & 3.70 & 0.952 & 7.91                  \\
            \hline

            TF-CrossNet   & 19.2 & 20.2    & 3.74 & 0.953 & 8.13                  \\

           \specialrule{1pt}{0pt}{0pt}
        \end{tabular}}

\end{table}

\subsubsection{Task-Standard Evaluations}
In Table~\ref{tab:smswsj_comparison_1ch} and Table~\ref{tab:smswsj_comparison_6ch}, we compare the task-standard evaluation of TF-CrossNet with other baseline systems on the SMS-WSJ corpus in 1-ch and 6-ch conditions, in standard speech separation metrics of SDR, SI-SDR, perceptual evaluation of speech quality (PESQ) \citep{rix2001pesq}, and extended short-time objective intelligibility (eSTOI) \citep{taal2011algorithm}, as well as WER. TF-CrossNet outperforms all baseline systems. Related separation results can also be found in \citet{vahid2024crossnet}.

\begin{table}[!t]
    \caption{Speaker separation results on SMS-WSJ (6-channel).}
    \label{tab:smswsj_comparison_6ch}
    \centering
    \resizebox{0.9\linewidth}{!}{
        \begin{tabular}{@{}lccccc@{}}
            \specialrule{1pt}{0pt}{0pt}
            Model    & SI-SDR   & SDR   & PESQ & eSTOI & WER  \\
            \hline
            Unprocessed    & -5.5   & -0.4   & 1.50 & 0.441 & 79.11  \\
            Oracle direct-path   & $\infty$ & $\infty$ & 4.50 & 1.000 & 6.16 \\
            \hline
            FasNet+TAC \citep{luo2020end}    & 8.6  & - & 2.37 & 0.771 & 29.80                 \\
            MC-ConvTasNet \citep{zhang2020end}             & 10.8     & -        & 2.78 & 0.844 & 23.10                 \\
            MISO$_1$ \citep{wang2021multi}                 & 10.2     & -        & 3.05 & 0.859 & 14.0                  \\
            LBT \citep{taherian2022multi}                  & 13.2     & 14.8     & 3.33 & 0.910 & 9.60                  \\
            MISO$_1$-BF-MISO$_3$ \citep{wang2021multi}     & 15.6     & -        & 3.76 & 0.942 & 8.30                  \\
            TF-GridNet (1-stage) \citep{tfgridnet_2023_wang} & 19.9     & 21.2     & 3.89 & 0.966 & 6.92                  \\
            TF-GridNet (2-stage) \citep{tfgridnet_2023_wang} & 22.8     & 24.9     & 4.08 & 0.980 & 6.76                  \\
            SpatialNet \citep{quan2023spatialnet}          & 25.1     & 27.1     & 4.08 & 0.980 & 6.70                  \\
            \hline

            TF-CrossNet   & 25.8     & 27.6    & 4.20 &  0.987 & 6.30                  \\

           \specialrule{1pt}{0pt}{0pt}
        \end{tabular}}

\end{table}

\subsubsection{Comparison of Different AMs}\label{sec:smswsj_comparison_backends}

In Table~\ref{tab:smswsj_diferent_am}, we compare the ASR performance among AMs trained on different data. The reverberant-noisy speech is denoted as \emph{reverb-noisy} and \emph{mixture} denotes unprocessed two-talker mixtures. In the penultimate row, given the 1-ch TF-CrossNet output, the reverb-noisy AM produces a 8.13\% WER, and if the training data is switched to direct-path speech, the WER gets lowered to 7.60\%, outperforming the previous best of 7.91\% \citep{tfgridnet_2023_wang} with two-stage training. For 6-ch TF-CrossNet, the switch from reverb-noisy to direct-path AM lowers the WER from 6.30\% to 5.74\%, outperforming the previous best \citep{quan2023spatialnet} by $14.3\%$ relatively. This switch elevates the ASR performance with only a third of training utterances. It demonstrates that with a strong separation frontend, the backend does not have to be trained on noisy speech. The mismatch between frontend output and backend noisy training data degrades the recognition performance of the mainstream approach. 

\begin{table}[!htbp]
    \centering
    \caption{ASR ($\%$WER) results of different AMs on different test data on SMS-WSJ. ${\dagger}$ denotes performance upper bound for each condition.}
    \label{tab:smswsj_diferent_am}
    \centering
    \resizebox{\linewidth}{!}{
    \begin{tabular}[width=\linewidth]{@{}l  cccc c}
        \specialrule{1pt}{0pt}{4pt}
         \multirow{2}{*}{Test Data} & \multicolumn{5}{c}{AM Train Data} \\
        &Reverb-noisy  & WSJ &  Direct-path & TF-CrossNet (1-ch)$^{\dagger}$ & TF-CrossNet (6-ch)$^{\dagger}$ \\
        
         \midrule
         Mixture & 79.11 & 90.97 &  90.05 & 90.37 &90.65  \\
         \midrule
         Reverb-noisy &8.52   & 50.70 &  48.18& 46.61 & 49.03  \\
        WSJ  &6.45   & 5.15 & 5.26 & 5.46 &  5.19  \\
         Direct-path & 6.16   & 5.56 & 5.23 & 5.30& 5.26  \\
         \midrule
         TF-CrossNet (1-ch) &  8.13  & 7.86  &  7.60 & 6.26 & 6.74 \\
         TF-CrossNet (6-ch) & 6.30   & 5.94 &  5.74 & 5.48 & 5.49  \\
         \specialrule{1pt}{0pt}{0pt}
    \end{tabular}
    }
\end{table}

It is worth noting that the AM trained on TF-CrossNet separated training set achieves 6.26\% WER for 1-ch and 5.49\% for 6-ch. Although such matched training produces lower WERs, these backends depend on the data from pre-trained TF-CrossNets. When a better frontend is available, retraining these backends is required to achieve optimal performance, making these models less practical. The results in Table~\ref{tab:smswsj_diferent_am} show that training the ASR backend on clean speech to elevate recognition performance should be preferred to an ASR backend trained on noisy speech.

\subsection{Results on SMS-WSJ-Large}\label{sec:eval_smswsj_large}
We compare the systems trained or fine-tuned on SMS-WSJ-Large and the pre-trained models on SMS-WSJ to investigate the effects of frontend training data on the overall performance of decoupled systems, and to find out which ASR backend is most promising in certain conditions. For the ASR backends, we select the task-standard reverberant-noisy speech trained AM, and the direct-path speech trained AM as described in Section~\ref{sec:exp_backend_smswsj}.

\subsubsection{Evaluation in Single-channel Setup}\label{sec:results_smswsj_large_1ch}

\begin{table}[!htbp]
    \centering
    \caption{ASR ($\%$WER) comparison results on SMS-WSJ-Large (1-ch) test set. Each entry denotes the WER generated from reverb-noisy speech trained AM (left) and direct-path speech trained AM (right) separated by a slash.}
    \label{tab:smswsj_large_asr_1ch}
    \centering
    \resizebox{\linewidth}{!}{
    \begin{tabular}[width=\linewidth]{c c cccc}
        \specialrule{1pt}{0pt}{4pt}
         \multirow{2}{*}{System} & \multirow{2}{*}{T60} & \multicolumn{4}{c}{SNR}\\
         &  & 0-10 dB & 10-20 dB & 20-30 dB & 30-40 dB \\
         \midrule
         \multirow{3}{*}{\makecell[c]{SMS-WSJ\\Pre-trained}} & 0.2-0.5 s & 42.11/43.10 & 12.49/12.72 & 8.06/7.80 & 7.25/6.85  \\
         & 0.5-0.8 s & 51.80/54.98 & 16.15/17.77 & 9.89/10.44 & 8.94/9.14 \\
         & 0.8-1.1 s & 64.97/69.32 & 24.40/27.76 & 14.65/15.90 & 13.02/13.90 \\
         \midrule
         \multirow{3}{*}{SMS-WSJ-Large} & 0.2-0.5 s & 34.43/34.24 & 11.86/11.85 & 8.05/7.61 & 7.21/6.71  \\
         & 0.5-0.8 s & 40.54/41.40 & 14.12/14.70 & 9.17/9.59 & 8.38/8.48 \\
         & 0.8-1.1 s & 47.88/49.78 & 17.45/19.35 & 11.60/12.50 & 10.38/10.96 \\
         \specialrule{1pt}{0pt}{0pt}
    \end{tabular}
    }
\end{table}

The evaluation results on 1-ch SMS-WSJ-Large are presented in Table~\ref{tab:smswsj_large_asr_1ch}. We compare the separated speech by the TF-CrossNets fine-tuned on SMS-WSJ and on SMS-WSJ-Large, in three T60 and four SNR ranges. For each test condition and system, two WERs are computed based on the reverb-noisy speech trained and the direct-path speech trained AM. By comparing the WERs, we observe that, by increasing the size of the training dataset, WERs are lowered on both AMs. This meets the expectation that more training data translates to improved frontend performance, and reduces the processing artifacts in separated speech that degrade ASR backend performance.

\begin{figure}[t!]
    \centering
    \begin{subfigure}{0.48\textwidth}
        \centering
        \includegraphics[width=\linewidth]{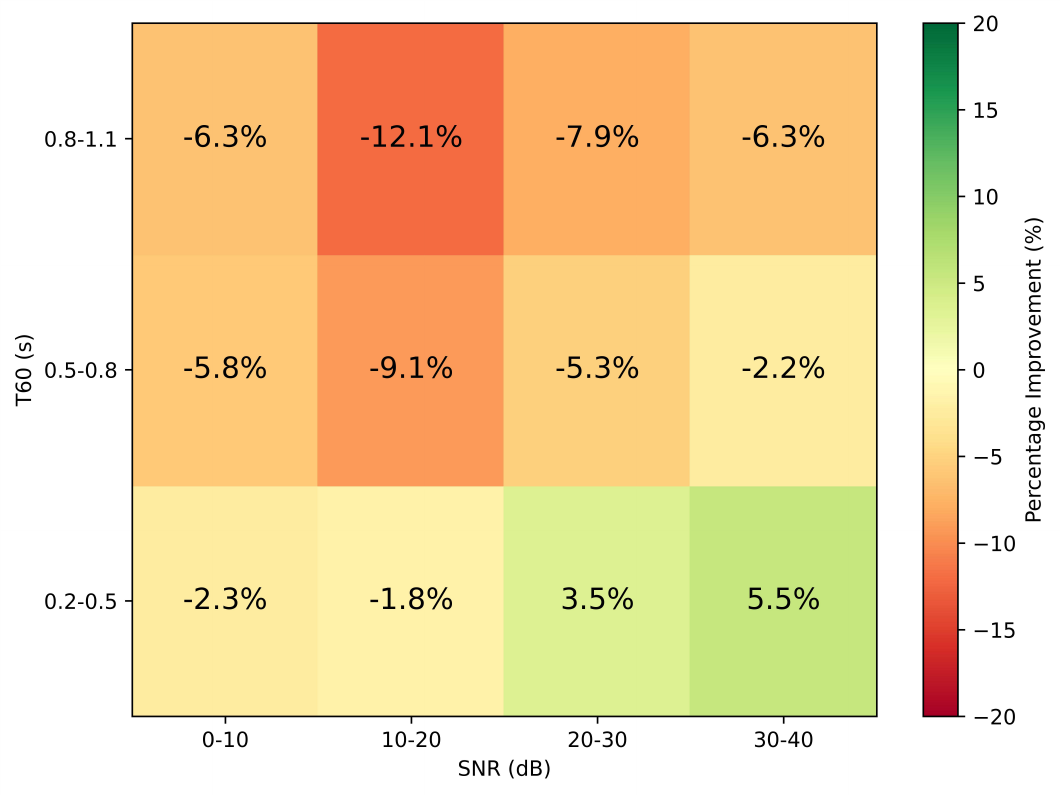}
        \caption{SMS-WSJ Pre-trained}
        \label{fig:smswsj_large_1ch_pretrained}
    \end{subfigure}
    \hfill
    \begin{subfigure}{0.48\textwidth}
        \centering
        \includegraphics[width=\linewidth]{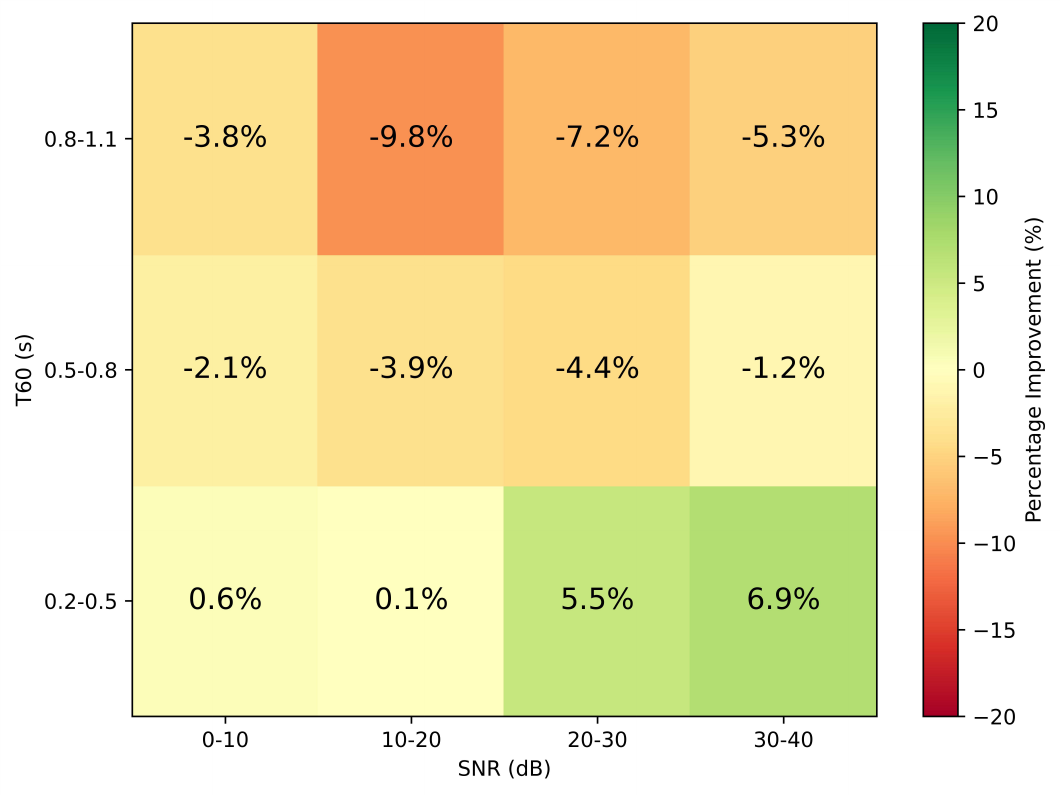}
        \caption{SMS-WSJ-Large Trained}
        \label{fig:smswsj_large_1ch_finetune}
    \end{subfigure}
    \caption{Visualization of the percentage improvement of the direct-path over reverb-noisy speech trained AM on SMS-WSJ-Large test set (1-ch). Negative numbers indicate ASR degradation.}
    \label{fig:smswsj_large_1ch}
\end{figure}

To better understand relative WERs in Table~\ref{tab:smswsj_large_asr_1ch}, we visualize the percentage improvement comparing the direct-path speech trained AM over the reverberant-noisy speech trained AM in Figure~\ref{fig:smswsj_large_1ch}. Each block denotes a test condition, and the number in each block denotes the relative WER reduction by switching the ASR backend from reverberant-noisy speech trained AM to direct-path speech trained AM. A color bar is used to denote the value of relative WER reduction. Positive numbers in green denote that the system produces lower WER with direct-path speech trained AM, and negative numbers in red denote that the system produces lower WER with reverberant-noisy speech trained AM. For example, the 5.5\% in green for 30-40 dB SNR and 0.2-0.5 s T60 denotes that the WER produced by the direct-path speech trained AM outperforms the reverberant-noisy speech trained AM by 5.5\%. For TF-CrossNet pre-trained on 1-ch SMS-WSJ, only two blocks are green in Figure~\ref{fig:smswsj_large_1ch_pretrained}, with maximum improvement of 5.5\%. For TF-CrossNet trained on 1-ch SMS-WSJ-Large, with more frontend training data, the blocks for all test conditions when the T60 range is 0.2-0.5 s are green, and the maximum improvement increases to 6.9\%. This finding indicates that with more frontend training data, decoupling the frontend and backend will more likely benefit robust multi-talker ASR. However, for 1-ch systems, when the test condition is highly adverse, say with low SNR or high T60, the mainstream approach to train ASR backend on noisy speech is still advantageous, and the direct-path speech trained AM elevates the overall performance in moderately and mildly adverse test conditions.

\subsubsection{Evaluation in Six-channel Setup}\label{sec:results_smswsj_large_6ch}
Table~\ref{tab:smswsj_large_asr_6ch} presents the WER comparisons this array setup. The improvements achieved by training TF-CrossNet on 6-ch SMS-WSJ-Large instead of SMS-WSJ are consistently and significantly larger compared with those in Table~\ref{tab:smswsj_large_asr_1ch}. On the task-standard test condition with T60 of 0.2-0.5 s and SNR of 20-30 dB, the WER with direct-path speech trained AM is lowered to 5.63\% compared with the 5.74\% result in Section~\ref{sec:smswsj_comparison_backends}.

\begin{table}[!b]
    \centering
    \caption{ASR ($\%$WER) comparison results on SMS-WSJ-Large (6-ch) test set. Each entry denotes the WER generated from reverb-noisy speech trained AM (left) and direct-path speech trained AM (right) separated by a slash.}
    \label{tab:smswsj_large_asr_6ch}
    \centering
    \resizebox{\linewidth}{!}{
    \begin{tabular}[width=\linewidth]{c c cccc}
        \specialrule{1pt}{0pt}{4pt}
         \multirow{2}{*}{System} & \multirow{2}{*}{T60} & \multicolumn{4}{c}{SNR}\\
         &  & 0-10 dB & 10-20 dB & 20-30 dB & 30-40 dB \\
         \midrule
         \multirow{3}{*}{\makecell[c]{SMS-WSJ\\Pre-trained}} & 0.2-0.5 s & 21.92/24.51 & 7.97/8.05 & 6.38/5.74 & 6.19/5.41  \\
         & 0.5-0.8 s & 29.25/34.53 & 8.77/9.46 & 6.59/6.18  & 6.28/5.75 \\
         & 0.8-1.1 s & 40.47/48.31 & 10.06/11.90 & 7.13/6.97 & 6.92/6.79 \\
         \midrule
         \multirow{3}{*}{SMS-WSJ-Large} & 0.2-0.5 s & 13.98/14.32 & 7.51/6.76 & 6.45/5.63 & 6.28/5.42 \\
         & 0.5-0.8 s & 16.12/16.83 & 7.72/7.24 & 6.56/5.84 & 6.30/5.55  \\
         & 0.8-1.1 s & 18.86/20.58 & 8.28/8.14 & 6.83/6.22 & 6.46/5.85 \\
         \specialrule{1pt}{0pt}{0pt}
    \end{tabular}
    }
\end{table}

\begin{figure}[!htbp]
    \centering
    \begin{subfigure}{0.48\textwidth}
        \centering
        \includegraphics[width=\linewidth]{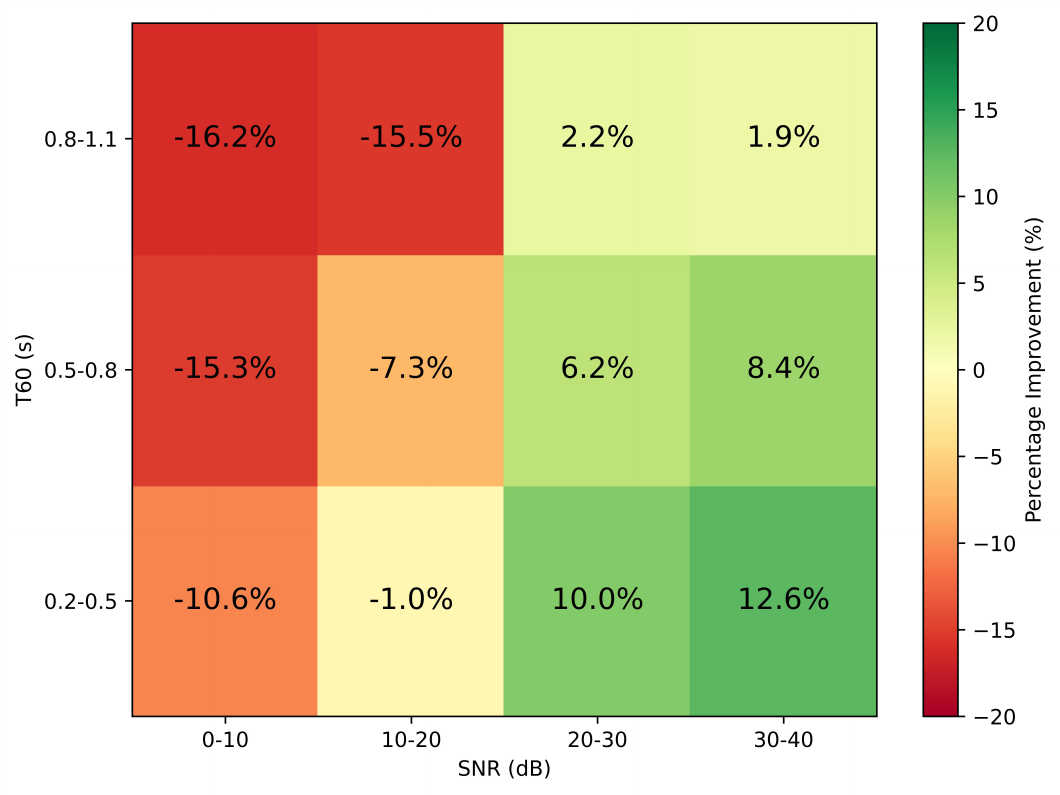}
        \caption{SMS-WSJ Pre-trained}
        \label{fig:smswsj_large_6ch_pretrained}
    \end{subfigure}
    \hfill
    \begin{subfigure}{0.48\textwidth}
        \centering
        \includegraphics[width=\linewidth]{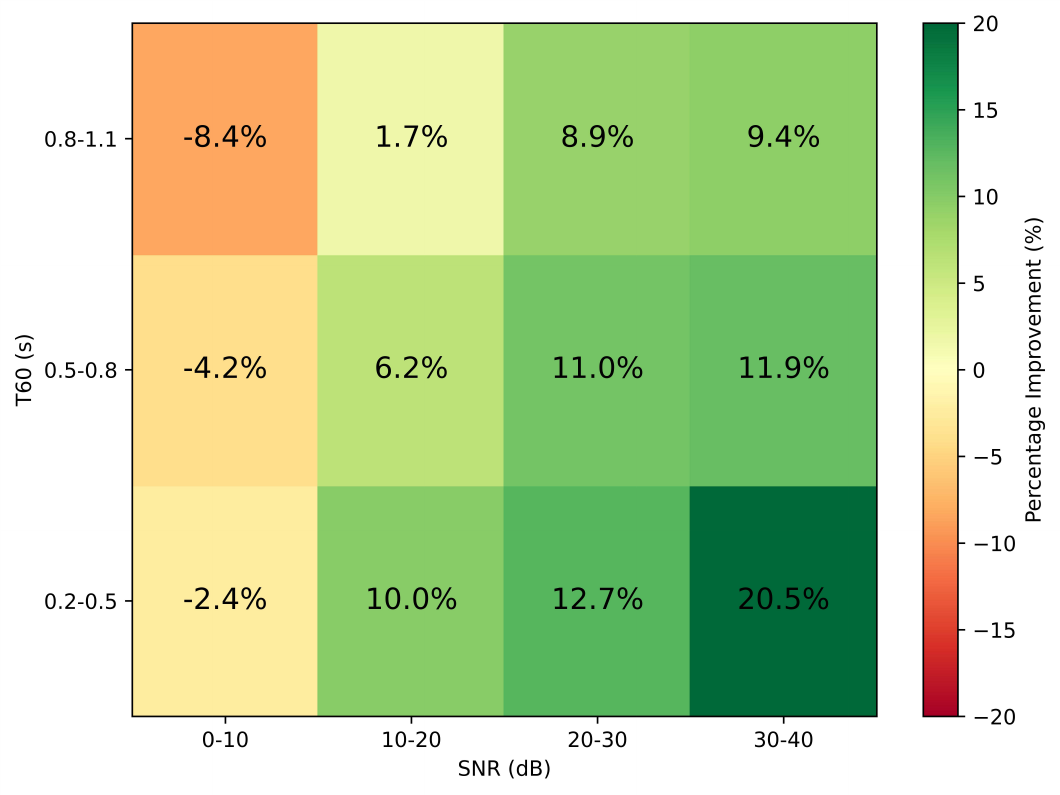}
        \caption{SMS-WSJ-Large Trained}
        \label{fig:smswsj_large_6ch_finetune}
    \end{subfigure}
    \caption{Visualization of the percentage improvement of the direct-path over reverb-noisy speech trained AM on SMS-WSJ-Large test set (6-ch).}
    \label{fig:smswsj_large_6ch}
\end{figure}

Figure~\ref{fig:smswsj_large_6ch} visualizes the percentage improvement of the direct-path speech trained AM relative to the reverberant-noisy speech trained AM. With 6-ch inputs instead of 1-ch, the separation frontend leverages spatial information and has better separation quality compared with the 1-ch model. This leads to more green blocks in Figure~\ref{fig:smswsj_large_6ch} compared with Figure~\ref{fig:smswsj_large_1ch}. When the frontend is trained on SMS-WSJ, the direct-path trained AM produces lower WER relative to the reverberant-noisy speech trained AM for all test conditions when SNR is 20 dB or higher as shown in Figure~\ref{fig:smswsj_large_6ch_pretrained}. When the frontend training data is changed to SMS-WSJ-Large, all test conditions with SNR at 10 dB or higher show better performance with the direct-path speech trained AM as shown in Figure~\ref{fig:smswsj_large_6ch_finetune}. The largest improvement also increases from 12.6\% to 20.5\% by changing the frontend training data from SMS-WSJ to SMS-WSJ-Large.

\subsubsection{When to Decouple Separation and Recognition?}
The results in Section~\ref{sec:results_smswsj_large_1ch} and Section~\ref{sec:results_smswsj_large_6ch} provide insights into the acoustic conditions where the proposed decoupled approach performs better than the mainstream approach:

\begin{itemize}
    \item In the single-channel setup shown in Figure~\ref{fig:smswsj_large_1ch}, with the pre-trained frontend on SMS-WSJ, the decoupled approach outperforms the mainstream approach for T60 in 0.2-0.5 s and SNR in 20-40 dB. With the frontend trained on SMS-WSJ-Large, the decoupled approach performs better for T60 in 0.2-0.5 s and SNR in 0-40 dB. The mainstream approach achieves lower WER at higher T60 and lower SNR values.
    \item In the six-channel setup shown in Figure~\ref{fig:smswsj_large_6ch}, with T60 in 0.2-1.1 s and SNR in 20-40 dB, the decoupled approach with the pre-trained frontend outperforms the mainstream approach. When the frontend is trained on SMS-WSJ-Large, the decoupled approach performs better for T60 in 0.2-1.1 s and SNR in 10-40 dB.
    \item With the same test SNR, T60, and training data, there are more conditions in the six-channel setup than in the single-channel setup under which the decoupled approach performs better. Furthermore, under the same conditions, the six-channel setup obtains higher amounts of improvement than the single-channel setup. We attribute this to better speaker separation of the multi-channel system relative to the single-channel system \citep{vahid2024crossnet}.
\end{itemize}

These findings suggest that decoupling the frontend and backend is an effective approach when the separation frontend produces a quality output. Such an output can be obtained at relatively low T60s and high SNRs, or with a powerful separation frontend.

\subsection{Results on LibriCSS}

\begin{table*}[!b]
    \centering
    \caption{cpWER (in $\%$) results for different separation and diarization methods on LibriCSS.}
    \label{tab:libricss_ours}
    \centering
    \resizebox{0.99\linewidth}{!}{
    \begin{tabular}[width=0.99\linewidth]{ l c c ccccccc}
        \specialrule{1pt}{0pt}{4pt}
        \multirow{2}{*}{Separation Method} &  \multirow{2}{*}{Diarization Method} &  \multirow{2}{*}{ASR}& \multicolumn{6}{c}{Overlap Ratio} & \multirow{2}{*}{Avg.}\\
          & & & 0S & 0L & 10$\%$ & 20$\%$ & 30$\%$ & 40$\%$ & \\
        \hline
         Unprocessed & Oracle & Our E2E & 3.80 & 3.61 & 9.60 & 16.67 & 24.97 & 34.29 & 17.08 \\
         MIMO-BF-MIMO~\citep{taherian2024leveraging} & Oracle & E2E & 3.45 & 3.87 & 3.15 & 4.46 & 5.35 & 5.76 & 4.44 \\
         SSND \citep{taherian2023ssnd} & Oracle & E2E & 4.04 & 3.97 & 3.37 & 3.54 & 4.51 & 4.66 & 4.04\\
         SSND \citep{taherian2023ssnd} & Oracle & E2E-SSL & 2.28 & 2.38 & 2.36 & 2.27 & 2.68 & 2.47 & 2.42\\
         SSND & Oracle & Our E2E & 3.62 & 3.49 & 3.40 & 3.56 & 4.19 & 4.11 & 3.77\\
        SSND  & Oracle & Our E2E-SSL & 2.27 & 2.48 & 2.22 & 2.17 & 2.53 & 2.48 & 2.36\\
        SSND & Oracle (w/ relaxation) & Our E2E & 2.47 & 2.38 & 2.31 & 2.48 & 3.12 & 3.15 & 2.69 \\
        SSND & Oracle (w/ relaxation) & Our E2E-SSL & 2.02 & 1.92 & 1.95& 2.00 & 2.31 & 2.47 & 2.13\\
        \midrule
        Unprocessed \citep{taherian2023ssnd} & X-vector + SC \citep{raj2021integration} & E2E & 13.95 & 12.20 & 20.12 & 29.64 & 35.06 & 41.81 & 27.01\\
        Unprocessed & X-vector + SC &  Our E2E & 11.59 & 11.26 & 19.01 & 26.68 & 32.24 & 39.48 & 24.83 \\
        Mask-based MVDR~\citep{chen2020libricss} & X-vector + SC & E2E & 8.78 & 13.07 & 10.51 & 15.37 & 17.54 & 17.63 & 14.13 \\
        MIMO-BF-MIMO~\citep{taherian2024leveraging} & X-vector + SC & E2E & 7.05 & 8.05 & 7.31 & 8.48 & 10.81 & 10.03 & 8.76  \\
        
        SSND \citep{taherian2023ssnd} & MC-EEND & E2E & 5.56 & 3.52 & 3.98 & 4.76 & 5.58 & 6.55 & 5.13 \\
         SSND & MC-EEND & Our E2E & 4.27 & 2.41 & 3.25 & 3.38 & 4.21 & 4.95 & 3.86 \\
         
        SSND & MC-EEND  & Our E2E-SSL & 3.43 & 2.77 & 2.15 & 2.29 & 3.06 & 3.69 & 2.92 \\
         \specialrule{1pt}{0pt}{0pt}
    \end{tabular}}
\end{table*}

We report the cpWER results of the proposed system on the LibriCSS corpus in Table~\ref{tab:libricss_ours}. On unprocessed multi-talker speech mixtures, with x-vector and spectral clustering (SC) diarization method \citep{raj2021integration}, the E2E ASR model achieves $27.01\%$ cpWER and our E2E model lowers it to $24.83\%$. With MC-EEND and SSND, we achieve $3.86\%$ cpWER with our E2E model, with only $0.09\%$ gap from the result with oracle diarization. Compared with the $1.09\%$ cpWER gap (from $5.13\%$ to $4.04\%$) in \citet{taherian2023ssnd}, our E2E model shows strong robustness to diarization errors as a backend for the CSS task. With WavLM-based feature extraction, we achieve 2.92\% cpWER with MC-EEND and SSND.

\begin{table}[!tbp]
    \centering
    \caption{Performance comparisons of speaker-attributed ASR systems on LibriCSS.}
    \label{tab:libricss_comparison}
    \centering
    \resizebox{\linewidth}{!}{
    \begin{tabular}[width=0.9\linewidth]{c c c c c}
        \specialrule{1pt}{0pt}{4pt}
        Ref. & Separation Method &  Diarization Method &  ASR& cpWER (\%)\\
        \hline
        \citet{wang2022localization} & CSS & DOA-based & TDNN-F \citep{raj2021integration} & 12.98\\
        \citet{raj2021integration} & CSS & X-vector + SC & E2E & 12.7\\
        \citet{kanda2022transcribe} & - & - & SA-ASR & 11.6\\
        \citet{delcroix2021speaker} & Speakerbeam & TS-VAD & E2E &18.8 \\
        \citet{delcroix2021speaker} & GSS & TS-VAD & E2E & 11.2 \\
        \citet{boeddeker2024ts} & \multicolumn{2}{c}{TS-SEP} & E2E & 6.42 \\
        \citet{boeddeker2024ts} & \multicolumn{2}{c}{TS-SEP} & E2E-SSL & 5.36 \\
        \citet{taherian2023ssnd} & \multicolumn{2}{c}{SSND} & E2E & 5.13\\
        \citet{taherian2023ssnd} & \multicolumn{2}{c}{SSND} & E2E-SSL & 3.22\\
        \hline
        Ours & \multicolumn{2}{c}{SSND} & Our E2E & 3.86 \\
        Ours & \multicolumn{2}{c}{SSND} & Our E2E-SSL & 2.92 \\

         \specialrule{1pt}{0pt}{0pt}
    \end{tabular}
    }
\end{table}

Further comparing the cpWER results of the proposed system with speaker diarization and the system with oracle utterance boundaries with our E2E model, we observe that the proposed system outperforms the oracle system in $0L$, $10\%$, $20\%$ overlap ratio conditions, which means that oracle decision boundaries can be further relaxed, since they may introduce extra insertion and deletion errors. We apply a relaxation collar -- 250 ms as traditionally chosen for speaker boundaries \citep{kalda2022collar} -- to both sides of oracle utterance boundaries for each speaker. With this relaxation collar, we achieve $2.69\%$ average cpWER, much lower than the $3.77\%$ result without relaxed boundaries. Boundary relaxation also reduces cpWER from 2.36\% to 2.13\% for our E2E-SSL model. This finding provides further room for progress towards oracle diarization.

A comprehensive comparison of the proposed system with other systems on speaker-attributed ASR is shown in Table~\ref{tab:libricss_comparison}. The E2E ASR model in the baselines is Transformer based and WRComformer is Conformer based. According to the ESPnet LibriCSS recipe \citep{watanabe2018espnet}, the Transformer-based ASR model outperforms the Conformer-based model. Our E2E system, based on WRConformer, successfully outperforms the previous best with the Transformer-based ASR model by $24.8\%$ relatively, upgrading the ESPnet findings. The $3.86\%$ cpWER obtained by our system represents the state-of-the-art result on LibriCSS with ASR backend trained on LibriSpeech, without leveraging SSL features. The results with our E2E-SSL outperform the 3.86\% result with our E2E model, and advance the state-of-the-art result 3.22\% in \citet{taherian2023ssnd} to 2.92\%. The results clearly demonstrate that separately improved ASR focusing on clean speech elevates the overall performance in a decoupled system.

\section{Concluding Remarks}\label{sec:conclusion}
We have proposed a decoupled approach to elevate the performance of robust multi-talker ASR. With powerful separation frontends available, an ASR backend trained on noisy speech may be suboptimal due to the mismatch between backend training data and separated speech. The proposed decoupled approach trains the frontend and backend separately, with the backend trained on clean speech only. The proposed approach is evaluated on several corpora. On Libri2Mix, we achieve a state-of-the-art 5.1\% WER by leveraging a speaker separation frontend with an ASR backend trained on Libri2Mix clean speech, outperforming the previous best by 27.1\%. On 1-ch and 6-ch SMS-WSJ, we achieve WERs of 7.60\% and 5.74\% respectively, outperforming the previous best systems trained on reverberant-noisy speech, and our backend is trained with a third of training utterances used in the baselines. On recorded LibriCSS, we elevate the ASR performance to a $2.92\%$ cpWER, outperforming the previous best by $9.3\%$. As a byproduct of the proposed approach, the capability of future frontend separation can be readily evaluated in terms of ASR by the backend acoustic model pre-trained on clean speech.

A further study on SMS-WSJ-Large suggests that the proposed decoupled approach is a strong alternative to the mainstream approach. In the single-channel setup, the decoupled approach shows better ASR performance than the mainstream approach with T60 in 0.2-0.5 s and SNR in 0-40 dB. In the six-channel setup, the decoupled approach performs better with T60 in 0.2-1.1 s and SNR in 10-40 dB. These findings show that the decoupled approach can elevate robust multi-talker ASR performance when the acoustic environment is not very reverberant and noisy, or a powerful separation algorithm is available. As speaker separation continues to advance, we expect that the decoupled approach becomes advantageous in more and more acoustic environments.

\section{Acknowledgement}
This work was supported in part by an NSF grant (ECCS-2125074), the Ohio Supercomputer Center, the NCSA Delta Supercomputer Center (OCI 2005572), and the Pittsburgh Supercomputer Center (NSF ACI-1928147).


 \bibliographystyle{elsarticle-harv} 
 \bibliography{yyf_refs}







\end{document}